\documentclass[prd,nofootinbib,showpacs,preprintnumbers,amsmath,amssymb]{revtex4}

\usepackage{graphicx}
\usepackage{dcolumn}
\usepackage{bm}

\def\bear{\begin{eqnarray}}
\def\ear{\end{eqnarray}}

\begin{document}

\title{Spin Resonance Effect on Pair Production in Rotating Electric Fields}

\author{Chul Min Kim}\email{chulmin@gist.ac.kr}
\affiliation{Center for Relativistic Laser Science, Institute for Basic Science, Gwangju 61005, Korea}
\affiliation{Advanced Photonics Research Institute, Gwangju Institute of Science and Technology, Gwangju 61005, Korea}

\author{Sang Pyo Kim}\email{sangkim@kunsan.ac.kr}
\affiliation{Department of Physics, Kunsan National University, Kunsan 54150, Korea}
\affiliation{Center for Relativistic Laser Science, Institute for Basic Science, Gwangju 61005, Korea}

\begin{abstract}
We advance a new analytical method for the Dirac equation in two-dimensional, homogeneous, time-dependent electric fields, which expresses the Cauchy problem of the two-component spinor and its derivative as the time-ordered integral of the transition rate of the time-dependent eigenspinors and the time-dependent energy eigenvalues. The in-vacuum at later times evolves from that at the past infinity and continuously make transitions between eigenspinors and between positive and negative frequencies of the time-dependent energy eigenvalues. The production of electron and positron pairs is given by the coefficient of the negative frequency at the future infinity which evolves from the positive frequency at the past infinity. In the adiabatic case when the time scale for the rotation of eigenspinors and energy eigenvalues is much longer than the electron Compton time, we find the spin-resonance effect on the pair production, which is simply determined by the spin rotation, the pair production and the continuous transmission coefficients between the energy eigenvalues without the change of spin states.
\end{abstract}

\date{\today}

\maketitle

\section{Introduction}\label{introduction}

Production of electron$-$positron pairs by a strong electric field predicted by Sauter and Schwinger is one of the most prominent nonperturbative phenomena in quantum electrodynamics (QED) \cite{sauter31,schwinger51}. The electromagnetic interaction of virtual electrons from the Dirac sea with the electric field transforms the in-vacuum into a different out-vacuum and results in the pair production. In the vector potential for the electric field, the positive frequency solution for the in-vacuum in the remote past scatters into one branch of positive frequency solution and another branch of negative frequency solution in the remote future. In the Coulomb potential the tunneling boundary condition defines the out-vacuum different from the in-vacuum. In a one-dimensional time-dependent vector potential the mode equation of the Dirac or the Klein-Gordon equation becomes a scattering problem. The pair production may be experimentally observed in extremely high intensity lasers \cite{DMHK12} or X-ray free electron lasers \cite{ringwald01} in the future. Thus, the quest of one-loop effective action and Schwinger pair production in an arbitrary electromagnetic field and their effects has been both theoretical and experimental issues.

The Dirac equation in a linearly polarized time-dependent electric field can be completely separated by eigenspinors and momenta into the spin-diagonal Fourier-component second-order equation \cite{blp-book}. The pair production is found from the Bogoliubov transformation between the in-vacuum and the out-vacuum. The exact solutions of the spin-diagonal equation, however, are known only for a few profiles of the electric fields \cite{bagrov-gitman90,dunne04,gelis-tanji}. As such, some methods have been proposed to approximately calculate the mean number for pair production in a general one-dimensional, time-dependent electric field. The worldline instanton method computes the instanton action along a Euclidean path in a given electric field, which gives the approximate pair-production rate \cite{dunne-schubert05,DWGS06}. The phase-integral and the adiabatic method find the pair-production rate from the coefficient of the negative frequency solution of the scattering matrix \cite{kim-page02,kim-page07,dumlu-dunne10,dumlu-dunne11,dabrowski-dunne14}. The Hamiltonian for a charged field in a time-dependent electric field can be extended to a complex plane and the quantum evolution of the in-vacuum along closed loops gives the pair-production rate in terms of the residues of the frequency \cite{kim13a,kim13b}. The branch cuts and poles for complex worldline instantons give pair production \cite{ITW1,ITW2}. The worldline dynamics is used to numerically compute the pair-production rate for general one-dimensional electric fields \cite{gies-klingmuller05}.

Another interesting problem in QED is pair production in multi-dimensional electromagnetic fields. The worldline instanton has been found for a specific configuration of two-dimensional electric fields \cite{dunne-wang06}. The split-operator formalism \cite{mocken-keitel04,mocken-keitel08,RMMHK09} and the Dirac-Heisenberg-Wigner formalism \cite{HAG10,HAG11} have been employed for numerical works for the Dirac equation in spacetime-dependent vector potentials. The inverse approach has been introduced to solve the Dirac equation in a specific configuration of spacetime-dependent electric fields \cite{oertel-schutzhold}. Pair production has been studied in a rotating magnetic field \cite{dipiazza-calucci02,kim14a}. Recently, pair production has been computed in a rotating time-dependent electric field \cite{blinne-gies14,strobel-xue14,strobel-xue15,blinne-strobel}. These rotating magnetic or electric fields are homogeneous and time-dependent. The circularly polarized field also gives a rotating electric and magnetic field on a fixed plane \cite{LLXSFL,WBK,LLX}. In strong contrast to the one-dimensional electric fields, however, the Dirac equation cannot be decomposed into eigenspinors and momenta since the eigenspinors explicitly depend on time due to the change of directions. In  perturbation theory, a time-dependent Hamiltonian induces continuous transitions of instantaneous eigenstates. In quantum cosmology, the Wheeler-DeWitt equation for the Friedmann-Robertson-Walker universe minimally coupled to a scalar field has a spectrum of intrinsic time-dependent eigenstates and induces the continuous transition among the eigenstates of the scalar field Hamiltonian \cite{kim92} (for references, see Ref. \cite{kim13c}). In scalar QED, the Klein-Gordon equation in a time-dependent magnetic field along a fixed direction has Landau states that continuously make transitions among different Landau states \cite{kim14a,kim14b}.

In this paper we introduce a new analytical method for the Dirac equation in a two-dimensional, homogeneous, time-dependent electric field and investigate the effect of spin resonance on pair production in a rotating electric field and the spin-changing effect on pair production in a general time-dependent electric field. The eigenspinors for electrons in a two-dimensional electric field with time-changing direction depend explicitly on time and thus make continuous transitions between different spin states. In fact, this results in the spin resonance in a rotating electric field and the spin-changing effect in a general electric field. The spin resonance or spin-changing effect is a characteristic feature of QED in a multi-dimensional electric field with time-changing direction, in strong contrast to the case of a fixed direction. Our stratagem is to write the second-order Dirac equation as two-eigenspinor component of first order equation, which has been introduced by Feshbach and Villars \cite{feshbach-villars58}. The two-component first order formalism has also been used for the Klein-Gordon equation in nontrivial spacetime backgrounds \cite{mostafazadeh06a,mostafazadeh06b}.

The continuous transition of eigenspinors in a direction-changing, time-dependent electric field is analogous to the continuous transition among Landau levels in a time-dependent magnetic field along a fixed direction. In the unidirectional, time-dependent magnetic field Landau levels continuously change in time, needless to say Landau energies. The rate of change of each Landau level to neighboring ones is determined by the derivative of the logarithm of the magnetic field, and the Cauchy initial value problem for the evolution of Landau states can be formulated in terms of the transition matrix of Landau levels and energies \cite{kim14a}. In the second quantized formulation of scalar QED, the Hamiltonian is equivalent to time-dependent oscillators with the time-dependent Landau energies, which are coupled by the coupling matrix \cite{kim14b}. In a similar manner, the rate of the direction change of a two-dimensional, time-dependent electric field determines the transition between eigenspinors. When the direction is fixed, the Dirac equation is completely separated by eigenspinors since the coupling between eigenspinors vanishes. In the case of a rotating electric field, the rate of the direction change is constant and leads to the spin resonance effect on quantum states and thereby pair production. We express the Cauchy problem for the two-eigenspinor component first-order equation of the Dirac equation by the time-ordered integral or Dyson series involving the eigenspinor coupling matrix and energy eigenvalues. In the case where the transition time between eigenspinors is much longer than the electron Compton time and the field intensity adiabatically changes, we obtain the leading term for pair production, which is given by the rate of changing direction, the pair production and the Wronskian between instantaneous energy eigenstates without the change of spin states.

The organization of this paper is as follows. In Sec. \ref{rotating field}, we formulate the Cauchy initial value problem for the second-order Dirac equation in a rotating electric field. For that purpose, we find the eigenspinors of the spin tensor in the rotating electric field and the rate of the change of eigenspinors (coupling matrix), and express the second-order equation as two-component first-order equation for spinors. Then the Cauchy problem is determined by the time-ordered integral or Dyson series of the scattering matrix of the energy eigenvalues and the coupling matrix.
In Sec. \ref{multi-E}, we extend the formulation to a homogeneous, time-dependent electric field with a changing direction. The spin coupling matrix comes from the rate of change of the field direction, the rotating electric field being the simplest one with a constant matrix. The Cauchy problem is similarly given by the time-ordered integral. In Sec. \ref{pair-prod}, we investigate the effect of spin resonance or spin change on the pair production of electrons and positrons. In the case of a slow change of eigenspinors in the electron Compton time, we explicitly obtain the leading term for pair production due to the spin resonance and time-dependent energy eigenvalues.

\section{Two-Component Spinor Formulation in Rotating Electric Field}\label{rotating field}

 A two-dimensional rotating electric field may be constructed by synthesizing two colliding laser pulses of equal intensities. If the two pulses have opposite circular polarizations, a pure rotating electric field is formed at every half a wavelength along the propagation axis. The charged particles located initially at such a locus will remain on the transverse plane during the interaction with the laser pulses. Such an experimental feasibility makes it meaningful to study to study pair production in a homogeneous, rotating electric field of the form
\begin{eqnarray}
{\bf E} (t) = (E(t) \cos (\Omega t), E(t) \sin (\Omega t), 0), \label{rot E}
\end{eqnarray}
with the vector potential ${\bf A} (t) = - \int_{- \infty}^t {\bf E}(t')dt'$.
Now, we advance a new analytical method that expresses the Dirac equation in the rotating electric field (\ref{rot E}) as the spin-diagonal two-component second-order equation.
The Dirac equation $(\gamma^{\mu} \hat{p}_{\mu} - m) \Psi = 0$ with the covariant derivative
$\hat{p}_{\mu } = i \partial_{\mu} - eA_{\mu}$ has solutions of the form
$\Psi = (\gamma^{\mu} \hat{p}_{\mu} + m) \Phi$, which  takes the second-order formulation (in units of $\hbar = c =1$) \cite{blp-book}
\begin{eqnarray}
\Biggl[\Bigl( \frac{\partial^2}{\partial t^2} + (i \nabla + e \vec{A} (t))^2 + m^2 \Bigr) I + i e \begin{pmatrix}
  0 & {\vec {\sigma}} \cdot {\bf E}(t) \\
{\vec{\sigma}} \cdot {\bf E}(t) & 0 \end{pmatrix} \Biggr] \Phi (t, \vec{x}) = 0, \label{2nd eq}
\end{eqnarray}
where the last term is the spin tensor and $\vec{\sigma}$ is the Pauli spin matrix.
The first set of eigenspinors $v_{\lambda}$ of the second-order equation (\ref{2nd eq}) with eigenvalues $i^{\lambda}$ is given by
\begin{eqnarray}
v_0 (t) =  \frac{1}{\sqrt{2}}\begin{pmatrix}
  e^{- i \Omega t/2} \\ 0 \\0 \\ e^{i \Omega t/2} \end{pmatrix}, \quad v_2 (t) =  \frac{1}{\sqrt{2}}\begin{pmatrix}
  e^{- i \Omega t/2} \\ 0 \\0 \\ - e^{i \Omega t/2} \end{pmatrix},   \label{eig 1}
\end{eqnarray}
and the second set with eigenvalues $i^{\lambda -1}$ by
\begin{eqnarray}
v_1 (t) =  \frac{1}{\sqrt{2}}\begin{pmatrix}
0\\  e^{i \Omega t/2} \\ e^{-i \Omega t/2} \\0 \end{pmatrix}, \quad v_3  (t) =  \frac{1}{\sqrt{2}}\begin{pmatrix}
0\\  e^{i \Omega t/2} \\ - e^{-i \Omega t/2} \\0 \end{pmatrix}.   \label{eig -1}
\end{eqnarray}
The factor $1/2$ in Eqs. (\ref{eig 1}) and (\ref{eig -1}) is a consequence of the spin structure, which implies that the eigenspinors return to the same value after a rotation of $4 \pi$ instead of $2 \pi$.

The eigenspinors (\ref{eig 1}) and (\ref{eig -1}) are orthonormal to each other
\begin{eqnarray}
v^{\dagger}_i v_j = \delta_{ij}. \label{orth}
\end{eqnarray}
The orthonormality (\ref{orth}) guarantees that the time derivative of the first set of eigenspinors is closed as
\begin{eqnarray}
\frac{d}{dt} \begin{pmatrix}
v_0 (t) \\  v_2 (t)  \end{pmatrix} = - i \frac{\Omega}{2} \sigma_1  \begin{pmatrix}
v_0 (t) \\  v_2 (t)  \end{pmatrix},
\end{eqnarray}
and that the second set is similarly closed
\begin{eqnarray}
\frac{d}{dt} \begin{pmatrix}
v_1 (t) \\  v_3 (t)  \end{pmatrix} =  i \frac{\Omega}{2} \sigma_1 \begin{pmatrix}
v_1 (t) \\  v_3 (t)  \end{pmatrix}.
\end{eqnarray}
Therefore, each eigenspinor set unitarily transforms the initial data as
\begin{eqnarray}
\begin{pmatrix} v_0 (t) \\  v_2 (t) \end{pmatrix} = e^{- i \frac{\Omega}{2} \sigma_1 (t - t_0)} \begin{pmatrix} v_0 (t_0) \\  v_2 (t_0) \end{pmatrix},
\end{eqnarray}
Here and hereafter, it is understood that matrix products act on the space of $\{ v_0, v_2 \}$ or $\{ v_1, v_3 \}$ separately but not on the spinor space itself. From now on, we shall work on the subspace of $v_0$ and $v_2$ since the eigenspinors are doubly degenerated. The eigenspinors periodically oscillate and lead to the spin resonance effect on pair production as will be shown in Sec. \ref{pair-prod}.
Interestingly, the change of eigenspinors is analogous to the change of Landau levels of a charged scalar in a homogeneous, time-dependent magnetic field along a fixed direction, in which the rate of the vector of all Landau levels is also given by a coupling matrix \cite{kim14a,kim14b}. It should be noted that the eigenspinors of an electron also change in a rotating magnetic field.

Using the unitary matrix for the rate of the change of two-component spinors
\begin{eqnarray}
S (t) =  e^{- i \frac{\Omega}{2} \sigma_1 (t-t_0)}, \label{tran mat1}
\end{eqnarray}
we may expand the spinor both by the eigenspinors and the Fourier component as
\begin{eqnarray}
\Phi (t, {\bf x}) = \int \frac{d^3 {\bf k}}{(2 \pi)^3} e^{i {\bf k} \cdot {\bf x}} {\bf v}^T (t) \cdot S^{\dagger} (t) \cdot {\bf \varphi}_{\bf k} (t), \label{spinor}
\end{eqnarray}
where $T$ denotes the transpose and we have used the compact notation
\begin{eqnarray}
{\bf {\varphi}}_{\bf k} (t) = \begin{pmatrix} \varphi_{0 {\bf k}} (t) \\  \varphi_{2 {\bf k}} (t)  \end{pmatrix}, \quad {\bf v} (t) = \begin{pmatrix} v_0 (t) \\  v_2 (t) \end{pmatrix}.
\end{eqnarray}
Then, the time derivative of the Fourier component of the spinor is given by
\begin{eqnarray}
\dot{\Phi}_{\bf k} (t) = {\bf v}^T (t) \cdot  S^{\dagger} (t) \cdot \dot{\bf \varphi}_{\bf k} (t).
\end{eqnarray}
The unitary matrix $S(t)$ has the effect of counter-rotating ${\bf v} (t)$ back to ${\bf v} (t_0)$, in a sense of the rotating frame with the electric field.

Substituting the spinor (\ref{spinor}) into Eq. (\ref{2nd eq}), we obtain the spin-diagonal Fourier-decomposed second-order equation
\begin{eqnarray}
\Bigl[\frac{d^2}{d t^2} + ({\bf k}_{\perp}-  e {\bf A}_{\perp} (t))^2 +k_z^2 +  m^2  + i e E(t)  S (t) \sigma_3 S^{\dagger} (t) \Bigr] \varphi_{\bf k} (t) = 0. \label{com eq}
\end{eqnarray}
Now, the second-order equation can be written as the two-component first-order equation
\begin{eqnarray}
\frac{d}{dt} \begin{pmatrix}
{\bf {\varphi}}_{\bf k} (t) \\  \dot{\bf{\varphi}}_{\bf k} (t)  \end{pmatrix} =  \begin{pmatrix}
 0 & I \\ - S (t) \pi_{\bf k}^2 (t) S^{\dagger} (t)  & 0 \end{pmatrix} \begin{pmatrix} {\bf {\varphi}}_{\bf k} (t) \\  \dot{\bf{\varphi}}_{\bf k} (t)  \end{pmatrix}, \label{2-com eq}
\end{eqnarray}
where
\begin{eqnarray}
\pi_{\bf k}^2 (t) = \bigl[({\bf k}_{\perp}-  e {\bf A}_{\perp} (t))^2 +k_z^2 +  m^2 \bigr]  I+ i e E (t) \sigma_3.
\end{eqnarray}
Following Ref. \cite{kim92} and using the similarity rule of the time-ordered integral \cite{dollard-friedman79}, the formal solution to the Cauchy problem for Eq. (\ref{2-com eq}) is given by the Dyson series, a time-ordered integral,\footnote{The time-ordered integral is the same as the product integral, ${\rm T} \exp \Bigl[\int_{t_0}^t O(t') dt \Bigr] = \prod_{t_0}^t \exp \Bigl[O(t') dt \Bigr]$ in Ref. \cite{dollard-friedman79}. Now, the similarity rule is that $\prod_{t_0}^t \exp \Bigl[O(t') dt \Bigr] = M (t) \prod_{t_0}^t \exp \Bigl[(M^{-1} (t') O(t') M(t') - M^{-1} (t') \dot{M} (t')) dt \Bigr]$ for an invertible operator $M(t)$. Putting $S \pi_{\bf k}^2 S^{\dagger}$ for $O$ and $S$ for $M$ gives Eq. (\ref{2-com sol}).}
\begin{eqnarray}
\begin{pmatrix}
{\bf {\varphi}}_{\bf k} (t) \\  \dot{\bf{\varphi}}_{\bf k} (t)   \end{pmatrix} =  S (t) \cdot {\rm T} \exp \Bigl[ \int_{t_0}^{t} \begin{pmatrix}
 i \frac{\Omega}{2} \sigma_1 & I \\ - \pi_{\bf k}^2 (t')   & i \frac{\Omega}{2} \sigma_1 \end{pmatrix} dt' \Bigr] \begin{pmatrix}
{\bf {\varphi}}_{\bf k} (t_0) \\  \dot{\bf{\varphi}}_{\bf k} (t_0)   \end{pmatrix}. \label{2-com sol}
\end{eqnarray}
Note that the matrix in the exponent does not commute at different times, so the time-ordered integral is not an ordinary one but a time-ordered one.  Finally, we obtain the solution to the Cauchy problem of two-component spinors
\begin{eqnarray}
\begin{pmatrix}
{\Phi}_{\bf k} (t) \\  \dot{\Phi}_{\bf k} (t)   \end{pmatrix} =  {\bf v}^{T} (t) \cdot  {\rm T} \exp \Bigl[ \int_{t_0}^{t} \begin{pmatrix}
 i \frac{\Omega}{2} \sigma_1 & I \\ - \pi_{\bf k}^2 (t')   & i \frac{\Omega}{2} \sigma_1 \end{pmatrix} dt' \Bigr] \begin{pmatrix}
{\bf {\varphi}}_{\bf k} (t_0) \\  \dot{\bf{\varphi}}_{\bf k} (t_0)   \end{pmatrix}. \label{fin sol}
\end{eqnarray}
Here, we have used the unitarity of $S^{\dagger} (t) S (t) = S (t) S^{\dagger} (t) = I$.

\section{Two-dimensional Time-Dependent Electric Field} \label{multi-E}

We now extend the rotating electric field in Sec. \ref{rotating field} to a general homogeneous, time-dependent electric field in a fixed plane
\begin{eqnarray}
{\bf E} (t) = (E_x(t), E_y(t), 0). \label{E}
\end{eqnarray}
Then, the spin tensor term in Eq. (\ref{2nd eq}) of the form
\begin{eqnarray}
\begin{pmatrix}
  0 & 0 & 0 & {\cal E}^* (t) \\
  0& 0& {\cal E} (t) &0\\
  0& {\cal E}^* (t)& 0 &0\\
  {\cal E} (t) &0&0&0 \end{pmatrix}  \label{spin gen}
\end{eqnarray}
with ${\cal E} (t) = E_x (t) + i E_y (t) = E(t) e^{i \theta (t)}$ has one set of eigenspinors with eigenvalues $i^{\lambda}$:
\begin{eqnarray}
v_0 (t) =  \frac{1}{\sqrt{2}}\begin{pmatrix}
  e^{- i \theta (t)/2} \\ 0 \\0 \\ e^{i \theta(t)/2} \end{pmatrix}, \quad v_2 (t) =  \frac{1}{\sqrt{2}}\begin{pmatrix}
  e^{- i \theta(t)/2} \\ 0 \\0 \\ - e^{i \theta(t)/2} \end{pmatrix}.   \label{mul 1}
\end{eqnarray}
Owing to the orthonormality $v^{\dagger}_i v_j = \delta_{ij}$, the subspace of $\{ v_0, v_2 \}$ changes as
\begin{eqnarray}
\frac{d}{dt} \begin{pmatrix}
v_0 (t) \\  v_2 (t)  \end{pmatrix} = - i \frac{\dot{\theta} (t)}{2} \sigma_1  \begin{pmatrix}
v_0 (t) \\  v_2 (t)  \end{pmatrix}.
\end{eqnarray}
Thus, the subspace unitarily transforms as
\begin{eqnarray}
\begin{pmatrix}
v_0 (t) \\  v_2 (t)  \end{pmatrix} = e^{- \frac{i}{2} (\theta (t) - \theta (t_0)) \sigma_1} \begin{pmatrix}
v_0 (t_0) \\  v_2 (t_0)  \end{pmatrix}.
\end{eqnarray}
Another set of degenerate eigenspinors with eigenvalues $i^{\lambda -1}$ is given by
\begin{eqnarray}
v_1 (t) =  \frac{1}{\sqrt{2}}\begin{pmatrix}
0\\  e^{i \theta(t)/2} \\ e^{-i \theta(t)/2} \\0 \end{pmatrix}, \quad v_3  (t) =  \frac{1}{\sqrt{2}}\begin{pmatrix}
0\\  e^{i \theta(t)/2} \\ - e^{-i \theta(t)/2} \\0 \end{pmatrix}   \label{mul 2}
\end{eqnarray}
and is closed in the subspace as
\begin{eqnarray}
\frac{d}{dt} \begin{pmatrix}
v_1 (t) \\  v_3 (t)  \end{pmatrix} = i \frac{\dot{\theta} (t)}{2} \sigma_1  \begin{pmatrix}
v_1 (t) \\  v_3 (t)  \end{pmatrix}.
\end{eqnarray}

Introducing the transition matrix for the eigenspinors
\begin{eqnarray}
S (t) = e^{- \frac{i}{2} (\theta (t) - \theta (t_0)) \sigma_1}, \label{tran mat2}
\end{eqnarray}
the spin-diagonal Fourier-component (\ref{2nd eq}) for the eigenspinors (\ref{mul 1}) now takes the form
\begin{eqnarray}
\Bigl[\frac{d^2}{d t^2} + ({\bf k}_{\perp}-  e {\bf A}_{\perp} (t))^2 +k_z^2 +  m^2  + i e E (t) S (t) \sigma_3 S^{\dagger} (t) \Bigr] \varphi_{\bf k} (t) = 0. \label{mul eq}
\end{eqnarray}
Following the case of the rotating electric field, we find the formal solution to Eq. (\ref{2-com eq}) as the time-ordered integral
\begin{eqnarray}
\begin{pmatrix}
{\bf {\varphi}}_{\bf k} (t) \\  \dot{\bf{\varphi}}_{\bf k} (t)   \end{pmatrix} =  S (t) \cdot {\rm T} \exp \Bigl[ \int_{t_0}^{t} \begin{pmatrix}
\frac{i}{2} \dot{\theta} (t') \sigma_1  & I \\ - \pi_{\bf k}^2 (t')  & \frac{i}{2} \dot{\theta} (t') \sigma_1  \end{pmatrix} dt' \Bigr] \begin{pmatrix}
{\bf {\varphi}}_{\bf k} (t_0) \\  \dot{\bf{\varphi}}_{\bf k} (t_0)   \end{pmatrix}. \label{2-com sol2}
\end{eqnarray}
where
\begin{eqnarray}
\pi_{\bf k}^2 (t) = \bigl[({\bf k}_{\perp}-  e {\bf A}_{\perp} (t))^2 +k_z^2 +  m^2 \bigr]  I+ i e E (t) \sigma_3.
\end{eqnarray}
Finally, we obtain the two-component spinor solution to the Dirac equation
\begin{eqnarray}
\begin{pmatrix}
{\Phi}_{\bf k} (t) \\  \dot{\Phi}_{\bf k} (t)   \end{pmatrix} =  {\bf v}^{T} (t) \cdot {\rm T} \exp \Bigl[ \int_{t_0}^{t} \begin{pmatrix}
\frac{i}{2} \dot{\theta} (t') \sigma_1 & I \\ - \pi_{\bf k}^2 (t')   & \frac{i}{2} \dot{\theta} (t') \sigma_1 \end{pmatrix} dt' \Bigr] \begin{pmatrix}
{\bf {\varphi}}_{\bf k} (t_0) \\  \dot{\bf{\varphi}}_{\bf k} (t_0)   \end{pmatrix}. \label{fin sol2}
\end{eqnarray}

\section{Pair Production}\label{pair-prod}

For simplicity reasons, we assume that the electric field is turned on for a finite period of time and choose the gauge ${\bf A} = 0$ at the past infinity and ${\bf A} = {\rm constant}$ at the future infinity. The Fourier-component spinor solution (\ref{fin sol2}) is entirely determined by the time-ordered integral
\begin{eqnarray}
{\cal U} (t, t_0) =  {\rm T} \exp \Bigl[ \int_{t_0}^{t} \begin{pmatrix}
\frac{i}{2} \dot{\theta} (t') \sigma_1 & I \\ - \pi_{\bf k}^2 (t')   & \frac{i}{2} \dot{\theta} (t')  \sigma_1 \end{pmatrix} dt' \Bigr]. \label{time int}
\end{eqnarray}
The rotating field is the specific case of $\dot{\theta} = \Omega$. Though Eq. (\ref{time int}) is valid for any profile of the electric field, we shall focus on the adiabatic case in which $|\dot{\theta} (t) | \ll m$, which means that the time scale for a significant change of the direction is much longer than the electron Compton time $t_C = 1/m$, which is the case for electric fields from lasers or X-ray free electron lasers. Then, we may separate the exponent into two parts
\begin{eqnarray}
{\cal M} (t) = {\cal M}_0 (t) + \delta {\cal M} (t) = \begin{pmatrix}
 0 & I \\ - \pi_{\bf k}^2 (t)   & 0 \end{pmatrix} +  \frac{i}{2} \dot{\theta} (t) \begin{pmatrix}
 \sigma_1 & 0 \\ 0   &  \sigma_1 \end{pmatrix}, \label{exp}
\end{eqnarray}
and apply the perturbation theory, in which the first off-diagonal matrix denoted by ${\cal M}_0$ is an unperturbed part and the second diagonal matrix denoted by $\delta {\cal M}$ is a perturbation.

The time-ordered integral for the matrix ${\cal M}_0$ can be expressed in terms of the solutions to the diagonal matrix equation
\begin{eqnarray}
\Bigl[ \frac{d^2}{dt^2} + \pi_{\bf k}^2 (t)  \Bigr] P_{\bf k} (t) = 0. \label{diag sol}
\end{eqnarray}
In fact, using two independent solutions to Eq. (\ref{diag sol}), we evaluate the time-ordered integral for ${\cal M}_0$ as
\begin{eqnarray}
{\cal U}_0 (t) = {\rm T} \exp \Bigl[ \int^{t} {\cal M}_0 (t') dt' \Bigr] = \begin{pmatrix}
 P_{\bf k}^{(+)} (t) & P_{\bf k}^{(-)} (t) \\ \dot{P}_{\bf k}^{(+)} (t)   & \dot{P}_{\bf k}^{(-)} (t) \end{pmatrix},
\end{eqnarray}
where $P_{\bf k}^{(+)} (t)$ and $P_{\bf k}^{(-)} (t)$ denote the diagonal matrix of positive and negative frequency solutions, respectively. We then use the property of time-ordered integral (\ref{time int}), which leads to \cite{kim92}
\begin{eqnarray}
{\cal U} (t, t_0) =   {\cal U}_0 (t) {\rm T} \exp \Bigl[ \int_{t_0}^{t} {\cal U}^{-1}_0 (t') \delta {\cal M} (t') {\cal U}_0 (t') dt' \Bigr] {\cal U}_0 (t_0). \label{prod int}
\end{eqnarray}
Here, the matrix elements of
\begin{eqnarray}
{\cal U}^{-1}_0 (t') \delta {\cal M} (t') {\cal U}_0 (t') =  i \frac{\dot{\theta}}{2} {\cal U}^{-1}_0 (\sigma_1 \otimes I) {\cal U}_0
\end{eqnarray}
read
\begin{eqnarray}
L_{11} &=& -  \frac{\dot{\theta}}{2} \begin{pmatrix}
0 & \dot{p}_{0 {\bf k}}^{(-)} p_{2 {\bf k}}^{(+)} - \dot{p}_{2 {\bf k}}^{(+)} p_{0 {\bf k}}^{(-)}\\ \dot{p}_{2 {\bf k}}^{(-)} p_{0 {\bf k}}^{(+)} - \dot{p}_{0 {\bf k}}^{(+)} p_{2 {\bf k}}^{(-)}  & 0 \end{pmatrix}, \\
L_{12} &=& -  \frac{\dot{\theta}}{2} \begin{pmatrix}
0 & \dot{p}_{0 {\bf k}}^{(-)} p_{2 {\bf k}}^{(-)} - \dot{p}_{2 {\bf k}}^{(-)} p_{0 {\bf k}}^{(-)}\\ \dot{p}_{2 {\bf k}}^{(-)} p_{0 {\bf k}}^{(-)} - \dot{p}_{0 {\bf k}}^{(-)} p_{2 {\bf k}}^{(-)}  & 0 \end{pmatrix}, \\
L_{21} &=& -  \frac{\dot{\theta}}{2} \begin{pmatrix}
0 & \dot{p}_{2 {\bf k}}^{(+)} p_{0 {\bf k}}^{(+)} - \dot{p}_{0 {\bf k}}^{(+)} p_{2 {\bf k}}^{(+)}\\ \dot{p}_{0 {\bf k}}^{(+)} p_{2 {\bf k}}^{(+)} - \dot{p}_{2 {\bf k}}^{(+)} p_{0 {\bf k}}^{(-)}  & 0 \end{pmatrix}, \\
L_{22} &=& - \frac{\dot{\theta}}{2} \begin{pmatrix}
0 & \dot{p}_{0 {\bf k}}^{(+)} p_{2 {\bf k}}^{(-)} - \dot{p}_{2 {\bf k}}^{(-)} p_{0 {\bf k}}^{(+)}\\ \dot{p}_{2 {\bf k}}^{(+)} p_{0 {\bf k}}^{(-)} - \dot{p}_{0 {\bf k}}^{(-)} p_{2 {\bf k}}^{(+)}  & 0 \end{pmatrix},
\end{eqnarray}
where $p_{\lambda {\bf k}}^{( \pm)}$ denote the positive and negative frequency solutions for the spinor $v_{\lambda}$, $(\lambda = 0, 2)$, respectively.

Now, the two-component spinor may be divided into two parts
\begin{eqnarray}
\begin{pmatrix} \Phi_{\bf k} (t) \\
\dot{\Phi}_{\bf k} (t) \end{pmatrix} = \begin{pmatrix} \Phi^{(0)}_{\bf k} (t) \\
\dot{\Phi}^{(0)}_{\bf k} (t) \end{pmatrix} +  \begin{pmatrix} \delta \Phi_{\bf k} (t) \\
\delta \dot{\Phi}_{\bf k} (t) \end{pmatrix},
\end{eqnarray}
where
\begin{eqnarray}
\begin{pmatrix}
\Phi^{(0)}_{\bf k} (t) \\
\dot{\Phi}^{(0)}_{\bf k} (t) \end{pmatrix}  = {\bf v}^{T} (t) \begin{pmatrix}
 P_{\bf k}^{(+)} (t) & P_{\bf k}^{(-)} (t) \\ \dot{P}_{\bf k}^{(+)} (t)   & \dot{P}_{\bf k}^{(-)} (t) \end{pmatrix} \begin{pmatrix}
\bar{\varphi }_{\bf k} (t_0) \\  \dot{\bar{\varphi}}_{\bf k} (t_0)   \end{pmatrix} \label{un sp}
\end{eqnarray}
is the two-component spinor solution in the instantaneous eigenspinors and
\begin{eqnarray}
\begin{pmatrix} \delta \Phi_{\bf k} (t) \\
\delta \dot{\Phi}_{\bf k} (t) \end{pmatrix} = {\bf v}^{T} (t) \begin{pmatrix}
 P_{\bf k}^{(+)} (t) & P_{\bf k}^{(-)} (t) \\ \dot{P}_{\bf k}^{(+)} (t)   & \dot{P}_{\bf k}^{(-)} (t) \end{pmatrix} \Biggl( {\rm T} \exp \Bigl[ \frac{i}{2}  \int_{t_0}^{t} \dot{\theta}(t') {\cal U}^{-1}_0 (t') (\sigma_1 \otimes I) {\cal U}_0 (t') dt' \Bigr] - I \Biggr) \begin{pmatrix}
\bar{\varphi}_{\bf k} (t_0) \\  \dot{\bar{\varphi}}_{\bf k} (t_0) \end{pmatrix} \label{del sp}
\end{eqnarray}
is the perturbation from the change of eigenspinors. Here, the Cauchy initial data are rearranged as
\begin{eqnarray}
\begin{pmatrix}
\bar{\varphi}_{\bf k} (t_0) \\  \dot{\bar{\varphi}}_{\bf k} (t_0)  \end{pmatrix} =   \begin{pmatrix}
 P_{\bf k}^{(+)} (t_0) & P_{\bf k}^{(-)} (t_0) \\ \dot{P}_{\bf k}^{(+)} (t_0)   & \dot{P}_{\bf k}^{(-)} (t_0) \end{pmatrix}^{-1} \begin{pmatrix}
{\varphi}_{\bf k} (t_0) \\  \dot{\varphi}_{\bf k} (t_0)   \end{pmatrix}.
\end{eqnarray}

Assuming the adiabaticity of $||\delta {\cal M} (t)|| \ll || {\cal M}_0 (t) ||$ for some suitable measure, we may expand the time-ordered integral in Eq. (\ref{del sp}) as the Dyson perturbation series \cite{BCOR09}
\begin{eqnarray}
{\rm T} \exp \Bigl[\int_{t_0}^t M(t') dt' \Bigr] = I + \int_{t_0}^t M(t') dt' +  \int_{t_0}^{t} dt' M(t') \int_{t_0}^{t'}  dt'' M(t'') + \cdots, \label{dyson}
\end{eqnarray}
and further assume the electron to be initially in the eigenspinor with $v_0 (t_0)$ of the Minkowski vacuum
\begin{eqnarray}
\bar{\varphi}_{0{\bf k}} (t_0)  = \begin{pmatrix}
1 \\  0  \end{pmatrix}, \quad
\dot{\bar{\varphi}}_{0{\bf k}} (t_0)  = \begin{pmatrix}
0 \\  0  \end{pmatrix}.
\end{eqnarray}
Then, the unperturbed part
\begin{eqnarray}
\Phi^{(0)}_{0{\bf k}} (t) =   P_{0 {\bf k}}^{(+)} (t) v_0 (t) \label{unper}
\end{eqnarray}
remains in the eigenspinor $v_0 (t)$ at a later time $t$ whereas the perturbation from the second term of the Dyson series (\ref{dyson})
\begin{eqnarray}
\delta \Phi^{(1)}_{0 {\bf k}} (t) &=&  - \frac{1}{2} \Biggl( P_{2 {\bf k}}^{(+)} (t)
   \int_{t_0}^{t} dt' \dot{\theta} (t') \bigl[ \dot{p}_{2 {\bf k}}^{(-)} (t') p_{0 {\bf k}}^{(+)} (t')
-  \dot{p}_{0 {\bf k}}^{(+)} (t') p_{2 {\bf k}}^{(-)} (t') \bigr] \nonumber\\
 && + P_{2 {\bf k}}^{(-)} (t) \int_{t_0}^{t} dt' \dot{\theta} (t') \bigl[ \dot{p}_{0 {\bf k}}^{(+)} (t') p_{2 {\bf k}}^{(+)} (t')
-  \dot{p}_{2 {\bf k}}^{(+)} (t') p_{0 {\bf k}}^{(+)} (t') \bigr]  \Biggr) v_2 (t) \label{reson}
\end{eqnarray}
makes a transition to the other eigenspinor $v_2 (t)$. And a similar expression holds for $\delta \Phi^{(1)}_{2 {\bf k}} (t)$ for the other initial eigenspinor
\begin{eqnarray}
\bar{\varphi}_{2{\bf k}} (t_0)  = \begin{pmatrix}
0 \\  1  \end{pmatrix}, \quad
\dot{\bar{\varphi}}_{2{\bf k}} (t_0)  = \begin{pmatrix}
0 \\  0  \end{pmatrix}.
\end{eqnarray}

First, in the case of a fixed direction such that $\dot{\theta} = 0$, the positive and negative frequency solutions (\ref{un sp}) at the future infinity are connected to those at the past infinity through the Bogoliubov transformation as
\begin{eqnarray}
P_{\lambda {\bf k}}^{(+) {\rm in}} (t) &=& \alpha_{\lambda {\bf k}} P_{\lambda {\bf k}}^{(+){\rm out}} (t) + \beta_{\lambda {\bf k}} P_{\lambda {\bf k}}^{(-){\rm out}} (t), \\
P_{\lambda {\bf k}}^{(-) {\rm in}} (t) &=& \alpha^*_{\lambda {\bf k}} P_{\lambda {\bf k}}^{(-){\rm out}} (t) + \beta^*_{\lambda {\bf k}} P_{\lambda {\bf k}}^{(+){\rm out}} (t),
\end{eqnarray}
where the Bogoliubov relation holds
\begin{eqnarray}
\vert \alpha_{0 {\bf k}} \alpha_{2 {\bf k}} \vert + \vert \beta_{0 {\bf k}} \beta_{2 {\bf k}} \vert = 1.
\end{eqnarray}
Then, the mean number for pair production is given by
\begin{eqnarray}
N_{\bf k}^{(0)} = \vert \beta_{0 {\bf k}} \beta_{2 {\bf k}} \vert. \label{pair}
\end{eqnarray}
Second, in the case of a slowly changing direction, the mean number for pair production from the same state is given by Eq. (\ref{pair}), which is the leading term, whereas the mean number from the time-changing eigenspinors is given by
\begin{eqnarray}
\vert \beta^{\rm res}_{0{\bf k}} \vert^2 = \frac{1}{4} \Bigl\vert \beta_{2 {\bf k}}  \int_{- \infty}^{\infty} dt' \dot{\theta} (t') {\rm Wr} [p_{0 {\bf k}}^{(+)} (t'), p_{2 {\bf k}}^{(-)} (t')] +  \alpha^*_{2 {\bf k}} \int_{- \infty}^{\infty} dt' \dot{\theta} (t') {\rm Wr} [ p_{2 {\bf k}}^{(+)} (t'), p_{0 {\bf k}}^{(+)} (t')]  \Bigr\vert^2, \label{res pair}
\end{eqnarray}
where ${\rm Wr}$ denotes the Wronskian. For an adiabatically changing field $(|\dot{\pi}_{\lambda {\bf k}}| \ll |\pi_{\lambda {\bf k}}| )$, we may use the adiabatic solutions
\begin{eqnarray}
P_{\lambda {\bf k}}^{(\pm)} (t) = \frac{1}{\sqrt{2 \pi_{\lambda {\bf k}} (t)}} e^{\mp i \int^{t}_{t_0} \pi_{\lambda {\bf k}} (t') dt'},
\end{eqnarray}
where
\begin{eqnarray}
\pi_{0{\bf k}}^2 (t) &=& ({\bf k}_{\perp}-  e {\bf A}_{\perp} (t))^2 +k_z^2 +  m^2 + i e E (t), \nonumber\\
\pi_{2 {\bf k}}^2 (t) &=& ({\bf k}_{\perp}-  e {\bf A}_{\perp} (t))^2 +k_z^2 +  m^2 -  i e E (t).
\end{eqnarray}
The second integral of Eq. (\ref{res pair}) is suppressed due to rapid oscillations in comparison to slow oscillations of the first integral. Finally, for the rotating electric field, we obtain the pair-production formula from the spin resonance effect
\begin{eqnarray}
N^{\rm res}_{\bf k} = \Bigl(\frac{\Omega}{2} \Bigr)^2 N_{\bf k}^{(0)} \Bigl\vert \int_{- \infty}^{\infty} dt' \frac{1}{2} \Bigl(\sqrt{\frac{\pi_{2{\bf k}} (t')}{\pi_{0{\bf k}} (t')}} + \sqrt{\frac{\pi_{0{\bf k}} (t')}{\pi_{2{\bf k}} (t')}} \Bigr) e^{i (\pi_{2{\bf k}} (t') - \pi_{0{\bf k}} (t')) } \Bigr\vert^2, \label{sp res ef}
\end{eqnarray}
where we have included two eigenspinors for virtual electrons from the Dirac sea. The meaning of the integrand is the transmission coefficient for the positive frequency solution under a sudden change of $\pi_{0{\bf k}} (t)$ to $\pi_{2{\bf k}} (t)$ and vice versa at each moment. Remarkably, the pair production from the spin resonance is determined by the rotation of eigenspinors, and the pair production and continuous transmissions between energy eigenvalues without any spin change.

\section{Discussion and Conclusion} \label{conclusions}

The quest of pair production in multi-dimensional electromagnetic fields has been for long time one of the challenging theoretical problems in QED. Surprisingly, the Dirac or Klein-Gordon equation cannot be completely separated into the mode equations of momenta and eigenspinors even in two-dimensional, homogeneous, time-dependent electromagnetic fields. In a unidirectional, time-dependent magnetic field, for instance, the charged scalar has time-dependent Landau levels,  which induce continuous transitions among themselves through out the evolution. The Dirac equation in a unidirectional, time-dependent electric field completely separates into the spin-diagonal second-order mode equations with time-dependent energies, in analogy to nonrelativistic problem in space-dependent potentials. When the direction of the time-dependent electric field changes, however, the eigenspinors of the Dirac equation depend on time in a similar manner as Landau levels in the unidirectional, time-dependent magnetic field. Therefore, the eigenspinors cannot simply diagonalize the Dirac equation. The split-operator formalism and the Dirac-Heisenberg-Wigner formalism, though useful for numerical works, do not give an analytical expression for pair production in a multi-dimensional, time-dependent electric field.

In this paper we have advanced a new analytical method for the Dirac equation in  a two-dimensional, time-dependent, rotating electric field and a general two-dimensional, time-dependent electric field. The eigenspinors explicitly depend on time and the rate of change of eigenspinors induces continuous transitions between eigenspinors through the evolution. In particular, a rotating electric field results in spin resonance between eigenspinors at a constant transition rate. Following the Cauchy problem for a charged scalar in a unidirectional, time-dependent magnetic field, we have decomposed the Dirac equation by the time-dependent eigenspinors in the two-dimensional, time-dependent electric field and formulate the Cauchy problem for the two-component spinor and its first derivative by introducing the coupling matrix given by the rate of change of eigenspinors. We then have solved the Cauchy initial value problem by expressing the formal solution in terms of the time-ordered integral of the energy eigenvalues and the coupling matrix for eigenspinors.

The Cauchy problem gives the in-vacuum state at the future infinity, which evolves from the in-vacuum at the past infinity, and thereby the pair production. When the eigenspinors and energy eigenvalues vary slowly in the electron Compton time, which is the case of lasers and X-ray free electron lasers, the time-ordered integral can be expanded in Dyson series and the first term is the leading correction to pair production due to the spin change. The leading term is the pair production from the change of energy eigenvalues without spin change. In the case of a rotating electric field, the spin resonance has an effect on pair production, which is proportional to the square of the half of spin rotation as well as the pair production and continuous transmissions between two energy eigenvalues without spin change. The pair production under rotating electric field can be rendered experimentally observable by using two colliding laser pulses of equal intensities but opposite circularly polarizations.

The issue not pursued in this paper is the Dirac theory in rotating or multi-dimensional, homogeneous magnetic fields and spacetime-dependent electric fields. In a rotating magnetic field both the eigenspinors and Landau levels depend explicitly on time, whose rates of change induce the resonance both of eigenspinors and Landau levels. The formulation in this paper can thus be extended to the rotating or direction-changing magnetic fields. On the other hand, in a spacetime-dependent electric field, each Fourier mode continuously couples to other modes due to spatial inhomogeneity in addition to continuous transitions and couplings of eigenspinors in time. The Dirac theory in these more general electromagnetic fields is beyond the scope of this paper and will be addressed in future publication. The other issue not treated in this paper is the vacuum polarization in two-dimensional, homogeneous time-dependent electric fields. The in-vacuum at the future infinity evolves from the in-vacuum at the past infinity while the out-vacuum is defined by the positive frequency state of the Dirac equation at the future infinity. Then the scattering matrix between the out-vacuum and the in-vacuum gives the complex one-loop effective action, whose vacuum persistence amplitude is related to the pair production computed in this paper.

\acknowledgements
The authors would like to thank Alexander Blinne, Holger Gies and Eckhard Strobel for useful discussions during Conference on Extremely High Intensity Laser Physics (ExHILP), Max-Planck-Institut f\"{u}r Kernphysik, July 21-22, 2015 and also thank participants of LPHYS'16, Shanghai, August 21-25, 2015 for useful discussions. S.P.K. would like to thank W-Y. Pauchy Hwang for the warm hospitality at National Taiwan University, where this paper was completed. This work was supported by IBS (Institute for Basic Science) under IBS-R012-D1.

\end{document}